\documentstyle[12pt]{article}
\include{FEYNMAN}
\include{epsf}
\begin{document}
\begin{flushright}
GUTPA/98/10/05
\end{flushright}
\vskip .1in
\newcommand{\lapprox}{\raisebox{-0.5ex}{$\ 
\stackrel{\textstyle<}{\textstyle\sim}\ $}}
\newcommand{\gapprox}{\raisebox{-0.5ex}{$\ 
\stackrel{\textstyle>}{\textstyle\sim}\ $}}
\newcommand{\be}{\begin{equation}}
\newcommand{\ee}{\end{equation}}
\newcommand{\bea}{\begin{eqnarray}}
\newcommand{\eea}{\end{eqnarray}}
\newcommand{\vect}[2]{\left ( \begin{array}{c}
        #1 \\ #2 \end{array} \right )}
\newcommand{\diag}[1]{\mbox{diag}\left \{#1 \right \}}
\newcommand{\eV}{\, \mbox{eV}}
\newcommand{\GeV}{\, \mbox{GeV}}
\newcommand{\Tr}{\mbox{Tr}}
\newcommand{\vev}[1]{\langle #1 \rangle}

\begin{center}

{\Large \bf
Neutrino masses and mixings from an anomaly free
$SMG \times U(1)^2$ model.}

\vspace{50pt}

{\bf C.D. Froggatt and M. Gibson}

\vspace{6pt}

{ \em Department of Physics and Astronomy\\
 Glasgow University, Glasgow G12 8QQ,
Scotland\\}

\vspace{18pt}

{\bf H.B.  Nielsen}

\vspace{6pt}

{ \em The  Niels  Bohr Institute \\
Blegdamsvej 17, DK-2100 Copenhagen {\O}, Denmark \\}
\end{center}

\section*{ }
\begin{center}
{\large\bf Abstract}
\end{center}
A natural solution to the fermion mass hierarchy problem suggests
the existence of a partially conserved chiral symmetry. We show that
this can lead to a reasonably natural solution to the solar and
atmospheric neutrino problems without fine-tuning or the
addition of new low energy fermions. The atmospheric neutrino atmospheric
neutrino anomaly is given by large mixing between $\nu_{\mu}$
and $\nu_{\tau}$, with $\Delta m^2_{atm} \sim 10^{-3} \eV^2$, and
the solar neutrino deficit is due to nearly maximal electron
neutrino vacuum oscillations. We present an explicit model for
the neutrino masses which is an anomaly free Abelian extension of the
standard model that also yields a realistic charged fermion spectrum.

\thispagestyle{empty}
\newpage

\section{Introduction}

The observed hierarchy of charged fermion masses and quark mixing
angles strongly suggests the existence of an approximate chiral
flavour symmetry \cite{fn1} beyond the standard model (SM). 
In a previous paper \cite{natsol} we discussed
the implications of such a symmetry for neutrino masses and 
mixings. We showed that the most natural scenario would
correspond to nearly maximal mixing 
between $\nu_e$ and $\nu_{\mu}$ being responsible for both the
solar and atmospheric neutrino problems. However, the recent data on the
atmospheric neutrino zenith angle dependence from Super-Kamiokande
\cite{SK}
indicate that this solution no longer gives an acceptable fit to the
atmospheric neutrino data. In this paper we show that approximately conserved
chiral symmetries can still lead to a reasonably natural solution to the
solar and atmospheric neutrino problems, if we relax the assumptions we made
in \cite{natsol}. We shall also present an explicit model for the
neutrino masses and mixings, in which the chiral
flavour symmetry comes from an Abelian extension of the standard model
gauge group.

Previously we made two assumptions for the models
with approximately conserved chiral symmetries:
\begin{enumerate}
\renewcommand{\labelenumi}{\roman{enumi}.}
\item
The low energy fermion spectrum of the model is the same as
in the standard model -- in particular we have only three left-handed
neutrinos.
\item
The chiral symmetries lead to
elements of the effective light neutrino mass matrix 
$M_{\nu}$ which are of
different orders of magnitude,
apart from those elements which are equal due to the symmetry 
$M_{\nu} = M^T_{\nu}$.
\end{enumerate}
As we discussed in our earlier paper \cite{natsol}, the only natural solution
to the solar and atmospheric neutrino problems with these assumptions is if we
have nearly maximal $\nu_e - \nu_{\mu}$ mixing, and small mixing
with $\nu_{\tau}$. This no longer gives a good
description of the atmospheric neutrino data. We cannot obtain any
other types of solution as a direct consequence of the assumptions
(i) and (ii).

Assumption (i) implies that we must have a $3 \times 3$ symmetric
effective Majorana-like 
 neutrino mass matrix. With a hierarchy between the elements
of a symmetric mass matrix there are essentially two different
forms for the matrix depending on whether or not the diagonal 
elements dominate all of the eigenvalues. 
The first case leads to small mixing between
all three neutrinos, and this is unsuitable for a solution to the
atmospheric neutrino problem. The second case gives large mixing
between two nearly degenerate neutrinos, and small mixing
with the third (non-degenerate) neutrino. 
Since we have only three neutrinos we have two independent
mass-squared differences ($\Delta m^2_{ij}$) for the neutrinos. The
smaller of these $\Delta m^2$s determines the wavelength
of oscillation for the two largely mixed neutrinos, which we must
take to be $\nu_{\mu}$ and $\nu_{\tau}$ with $\Delta m^2_{23}
\sim 10^{-3} \eV^2$ (and consequently the
other mass-squared differences $\Delta m^2_{12} \sim \Delta m^2_{13}
> 10^{-3} \eV^2$) if we wish to explain all of the
data on the atmospheric neutrino problem. However, we cannot
also explain the solar neutrino problem, since the electron neutrino
is then only slightly mixed and the small angle MSW solution
requires $\Delta m^2 \sim 10^{-5} \eV^2$.
Hence we see that it is necessary to relax our assumptions.

The first assumption was made because of the desire for minimality
in our theory. We do not wish to introduce extra low energy fermions
unless it is absolutely necessary, and
consequently we will retain assumption (i) in this paper. 
The second
assumption is often satisfied in models with chiral (gauged) symmetry breaking;
however, it is not uncommon to find two order of magnitude equal
elements in the mass matrices. Indeed in the explicit model
(based on the anti-grand unified model (AGUT), \cite{AGUT,AGUT1}) from
our previous paper we found that
the (1, 1) and (2, 2) elements of the neutrino mass matrix were
approximately equal, although in that case this did not have
any effect on the phenomenology. Hence, in this paper
we shall relax the second assumption and consider the case
where there are two order of magnitude
equal elements in our mass matrix (other than those elements which are
exactly equal due to the symmetry of the mass matrix). We do not expect
these elements to be exactly equal, since that would generally require
fine-tuning which we are careful to avoid.

In the next section we discuss the structure of the neutrino
mass matrix we would expect to have for natural models of this type, and the
phenomenology of the neutrino oscillations. We will show that (with
no fine-tuning) we would typically obtain nearly maximal $\nu_e$
vacuum oscillations (with a linear combination of $\nu_{\mu} -\nu_{\tau}$)
for the solar neutrinos, and large
$\nu_{\mu} - \nu_{\tau}$ oscillations for atmospheric neutrinos. We
would expect to see nothing at LSND, much of the parameter space for which
has already been ruled out by Karmen \cite{karmen}, and Bugey \cite{bugey}.

Whilst there are numerous examples of models 
\cite{AGUT,AGUT1,Leurer,Ibanez,Ramond} which explain
the fermion spectrum using global $U(1)$ symmetries, or which
cancel gauged $U(1)$ anomalies using the supersymmetric Green-Schwarz
mechanism, it seems to have become a common belief 
\cite{Bijnens,Binetruy}
that it is not possible
to construct an anomaly free gauged Abelian extension of the SM which
yields a realistic fermion mass spectrum.
We present here an explicit anomaly free model with gauge group
$SMG \times U(1)^2$ (where $SMG$ is the SM gauge group), which
(with a non-minimal Higgs field spectrum) fits the charged
fermion mass spectrum and yields solutions to the solar
and atmospheric neutrino problems. 
 The charged fermion mass spectrum in this model is identical to
that predicted by the AGUT model.
However, the neutrino mass spectrum is considerably different
from that given by the AGUT, and we show in section \ref{sec:neutrino}
that it can yield neutrino masses of the form suggested in
section \ref{sec:phenom}. In order to obtain the required
neutrino spectrum, it is necessary to introduce an $SU(2)$ triplet
Higgs field with a suitable vacuum expectation value. We also discuss
some difficulty in naturally obtaining such a vacuum expectation
value for this Higgs field from the scalar potential. 

\section{Neutrino Phenomenology}
\label{sec:phenom}

In this section we shall examine the possible structures of the
effective $3 \times 3$ light 
neutrino mass matrix, which can arise in models with approximately
conserved chiral symmetries in a reasonably natural way. In the
following discussion we shall use the convention that
\bea
\Delta m^2_{ij} & = & | m^2_{\nu_i} - m^2_{\nu_j} |,\\
\Delta m^2_{12} & < & \Delta m^2_{23},
\eea
where $\nu_i$ is the $i$th neutrino mass eigenstate. We then require
$\Delta m^2_{23} \sim 10^{-3} \eV^2$ 
and large $\nu_{\mu} - \nu_{\tau}$ mixing for the atmospheric neutrinos.
We can have several types of solution
to the solar neutrino problem, such as  the well
known MSW and `just-so' solutions to the solar neutrino problem
with
\be
\Delta m^2_{solar} \sim 10^{-5}, \, 10^{-10} \eV^2
\ee
respectively. There is also some variation in the solar neutrino fluxes 
predicted by 
different solar models and this theoretical uncertainty
means that it is also possible to have an
`energy-independent' vacuum oscillation solution to the solar neutrino problem
\cite{conforto}. By `energy-independent' we mean that $\Delta m^2_{solar}$
is sufficiently large that many oscillation lengths lie between
the sun and the earth, and what we observe is the averaged flux
suppression which is the same for solar neutrinos of all energies.
Hence we can have
\be
10^{-10} \lapprox \Delta m^2_{12} = \Delta m^2_{solar} \lapprox 10^{-4} \eV^2,
\ee
where the upper limit comes from the constraint
that electron neutrino mixing does not make a large contribution
to the atmospheric neutrinos.
This type of solution does not agree well with the solar neutrino
data if we take both the experimental and theoretical solar neutrino
rates at face value. (The Bahcall-Pinsonneault (BP98) model \cite{BP98}
rules out this possibility at $99 \% \,C.L.$) However, we note that there
is still some freedom allowed in the choice of solar model. 

The analysis
of \cite{conforto} examines the possibility of having an energy-independent
solution if the true solar model lies somewhere within the range of currently
allowed solar models. Taking the energy-independent flux suppression ($F$) as a
free parameter they find
\be
F = 0.50 \pm 0.06
\ee
with a minimum $\chi^2$ of 8. If $F=0.5$ is not a free parameter (as in our
model below) then this
corresponds to a confidence level of $5 \%$.
Even if the BP98 solar model is correct, the requirement for an
energy-dependent solution to the solar neutrino problem rests essentially on
only one experiment (the Chlorine experiment.) Given the
possibility of unknown systematic errors we would prefer to avoid
relying too strongly on the result of any single experiment. Hence, whilst
the MSW and `just-so' solutions to the solar neutrino problem are empirically
favoured we still consider the simpler energy-independent solution 
(with maximal mixing between two neutrinos) to be a viable solution.
The amount of mixing will be large for the vacuum oscillation solutions,
and may be either large or small for the MSW solutions.

As we saw in our previous paper if we have a completely
hierarchical mass matrix (with all independent elements
of different orders of magnitude), the only solution to the solar and
atmospheric neutrino problems is to have nearly maximal
$\nu_e - \nu_{\mu}$ mixing responsible for both, which seems to be no longer
compatible with the atmospheric neutrino data. Hence
we shall now look at the possible mass matrices with order
of magnitude degeneracies between the elements. One possibility
would be to have an order of magnitude degeneracy in the charged
lepton mass matrix, leading to large mixing coming from the charged sector.
It has been shown elsewhere in the literature \cite{Grossman,Ross,Pati} 
that this can
yield an acceptable phenomenology, and we do not consider it further
here. So we now consider order of magnitude equal elements
in the neutrino mass matrix. There are essentially three
types of matrix which could potentially yield an acceptable
phenomenology with a small number of approximately equal elements,
\be
\begin{array}{ccc}
 \mathrm{I} & \mathrm{II} & \mathrm{III} \\
 \left(\begin{array}{ccc}
A & \times & \times \\
\times & \times & A\\
\times & A & \times \\
\end{array}\right) & \,
 \left(\begin{array}{ccc}
\times & \times & \times\\
\times & A & B\\
\times & B & C\\
\end{array}\right) & \,
\left(\begin{array}{ccc}
\times & A & B\\
A & \times & \times\\
B & \times & \times\\
\end{array}\right) 
\end{array}
\ee
where $\times$ denotes small elements and in each case
$A \sim B \sim C$. We shall call these textures
I, II and III respectively.

From the form of texture I we see that this texture would require
the imposition of an exact flavour symmetry relating $(M_{\nu})_{11}$
to $(M_{\nu})_{23}$, for which we have
no good reason. Hence we will not use texture I. In order
to have a good phenomenology, type II would require $AC \sim B^2$,
which is not unlikely to occur by chance. However, it also requires
three order of magnitude equal elements in the neutrino mass matrix,
which we do not consider likely in most models with approximately
conserved chiral symmetries. Nevertheless, it has been
obtained in a supersymmetric extension of the standard model
with approximately conserved gauged chiral symmetries \cite{Ramond}.
Type III has only two approximately equal elements and, as we
shall see in section \ref{sec:neutrino}, can occur reasonably naturally in
a specific model. In fact type III has previously been considered in the
literature in \cite{Barbieri}, where the structure of the mass matrix
is assumed to be
due to a global $L_e - L_{\mu} - L_{\tau}$ symmetry. 
The fine-tuned case where $B = A$ 
corresponds to the popular `bi-maximal mixing'
solution to the neutrino problems \cite{bi-max,bi-max2}.
All of the textures (I, II, and III) examined here have previously been
discussed in \cite{Barbieri2} by three of the authors of \cite{Barbieri}.
However, they claim there that flavour symmetries which lead to textures
II and III also yield large mixing from the charged lepton mass matrix.
We do not find this to be the case here.

The mass matrix texture of type III has the eigenvalues:
\be
\pm \sqrt{A^2 + B^2}, 0
\ee
and can be diagonalised by the mixing matrix:
\bea
U_{\nu} & \sim & \left( \begin{array}{ccc}
1 & 0 & 0\\
0 & \cos \theta & - \sin \theta \\
0 & \sin \theta & \cos \theta \\
\end{array} \right)
\left( \begin{array}{ccc}
\frac{1}{\sqrt{2}} & -\frac{1}{\sqrt{2}} & 0\\
\frac{1}{\sqrt{2}} & \frac{1}{\sqrt{2}} & 0\\
0 & 0 & 1\\
\end{array} \right)\\
& = & \left( \begin{array}{ccc}
\frac{1}{\sqrt{2}} & -\frac{1}{\sqrt{2}} & 0\\
\frac{1}{\sqrt{2}} \cos \theta & \frac{1}{\sqrt{2}} \cos \theta &
        -\sin \theta\\
\frac{1}{\sqrt{2}} \sin \theta & \frac{1}{\sqrt{2}} \sin \theta &
        \cos \theta\\
\end{array} \right) \label{eq:Unu}
\eea
where 
\be
\tan \theta = \frac{B}{A}.
\ee
From the first row of eq.
\ref{eq:Unu} we can see that $\nu_e$ is maximally
mixed between $\nu_1$ and $\nu_2$, so that its mixing does not
contribute to the atmospheric neutrino anomaly, and there
will be no effect observable at Chooz \cite{Chooz}
since we take $\Delta m^2_{12}
< 10^{-4} \eV^2$. The atmospheric neutrino
anomaly will be entirely due to large $\nu_{\mu} - \nu_{\tau}$
mixing and, in order that the mixing be large enough, we need
$\sin^2 2\theta \gapprox 0.7$ ($95 \% \,C.L$) which requires
\be
\label{eq:b/a}
0.56 \lapprox \frac{B}{A} \lapprox 1.8.
\ee
So although $A$ and $B$ must be order of magnitude degenerate,
it is not necessary to do any fine tuning. The solar
neutrino problem is explained by vacuum oscillations, although
whether it is an `energy-independent' or a `just-so' solution
will depend on the small elements which we have neglected. 
It is not entirely clear which of these types of solution will
be more likely to occur in models with chiral symmetry breaking.
We note however that the elements of $M_{\nu}$ which contribute
to the $\Delta m^2_{12}$ have to be about 8 orders of magnitude
smaller than the large elements A and B for the `just-so'
solution.
The solar neutrino problem cannot be explained in this model 
by an MSW type solution, since the mixing
of the electron neutrino is too large for this type
of solution.

\section{Constructing an anomaly free $SMG \times U(1)^2$ model}
\label{sec:extension}

We now introduce an anomaly free Abelian extension of the SM
which we shall use in the next section to obtain a neutrino
mass spectrum of the form we have just discussed. This extension has the
gauge group
\be
SMG \times U(1)_{f1} \times U(1)_{f2}
\ee
and we have only the standard model fermion spectrum at low energies.
We shall break $U(1)_{f1}$ and $U(1)_{f2}$  with a non-minimal set
of three Higgs fields, which are required to give a realistic charged
fermion spectrum and which leave the $SMG$ unbroken. The $SMG$ will be broken
down to $SU(3) \times U(1)$ by the usual Weinberg-Salam Higgs field,
although this will now also carry charges under $U(1)_{f1}$ and
$U(1)_{f2}$.
We shall also introduce a further Higgs field to generate a realistic
spectrum of neutrino
masses in the next section.

\begin{figure}
\begin{picture}(40000,10000)
\THICKLINES

\drawline\fermion[\E\REG](5000,1500)[4000]
\drawarrow[\E\ATBASE](\pmidx,\pmidy)
\global\advance \pmidy by -2000
\put(\pmidx,\pmidy){$b_L$}

\drawline\fermion[\E\REG](9000,1500)[4000]
\drawarrow[\E\ATBASE](\pmidx,\pmidy)
\global\advance \pmidy by -2000
\put(\pmidx,\pmidy){$M_F$}

\drawline\fermion[\E\REG](13000,1500)[4000]
\drawarrow[\E\ATBASE](\pmidx,\pmidy)
\global\advance \pmidy by -2000
\put(\pmidx,\pmidy){$M_F$}

\drawline\fermion[\E\REG](17000,1500)[4000]
\drawarrow[\E\ATBASE](\pmidx,\pmidy)
\global\advance \pmidy by -2000
\put(\pmidx,\pmidy){$M_F$}

\drawline\fermion[\E\REG](21000,1500)[4000]
\drawarrow[\E\ATBASE](\pmidx,\pmidy)
\global\advance \pmidy by -2000
\put(\pmidx,\pmidy){$b_R$}

\drawline\scalar[\N\REG](9000,1500)[4]
\global\advance \pmidx by 500
\global\advance \pmidy by 1500
\put(\pmidx,\pmidy){$\phi_{WS}$}
\global\advance \scalarbackx by -530
\global\advance \scalarbacky by -530
\drawline\fermion[\NE\REG](\scalarbackx,\scalarbacky)[1500]
\global\advance \scalarbacky by 1060
\drawline\fermion[\SE\REG](\scalarbackx,\scalarbacky)[1500]

\drawline\scalar[\N\REG](13000,1500)[4]
\global\advance \pmidx by 500
\global\advance \pmidy by 1500
\put(\pmidx,\pmidy){$W$}
\global\advance \scalarbackx by -530
\global\advance \scalarbacky by -530
\drawline\fermion[\NE\REG](\scalarbackx,\scalarbacky)[1500]
\global\advance \scalarbacky by 1060
\drawline\fermion[\SE\REG](\scalarbackx,\scalarbacky)[1500]

\drawline\scalar[\N\REG](17000,1500)[4]
\global\advance \pmidx by 500
\global\advance \pmidy by 1500
\put(\pmidx,\pmidy){$\theta$}
\global\advance \scalarbackx by -530
\global\advance \scalarbacky by -530
\drawline\fermion[\NE\REG](\scalarbackx,\scalarbacky)[1500]
\global\advance \scalarbacky by 1060
\drawline\fermion[\SE\REG](\scalarbackx,\scalarbacky)[1500]
\drawline\scalar[\N\REG](21000,1500)[4]
\global\advance \pmidx by 500
\global\advance \pmidy by 1500
\put(\pmidx,\pmidy){$\theta$}
\global\advance \scalarbackx by -530
\global\advance \scalarbacky by -530
\drawline\fermion[\NE\REG](\scalarbackx,\scalarbacky)[1500]
\global\advance \scalarbacky by 1060
\drawline\fermion[\SE\REG](\scalarbackx,\scalarbacky)[1500]

\end{picture}
\vskip .3cm
\caption{Feynman diagram for bottom quark mass in the full theory.
The crosses indicate the couplings of the Higgs fields to the vacuum.}
%\label{MbFull}
\label{fig:fdiag}
\end{figure}
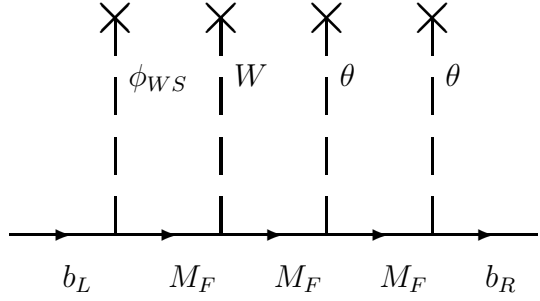

The fermions will each have different charges under the chiral symmetries
$U(1)_{f1}$ and $U(1)_{f2}$, which will prevent most of them from
acquiring masses by a direct Yukawa coupling with the Weinberg-Salam
Higgs field. However, after the spontaneous breaking of $U(1)_{f1}$ 
and $U(1)_{f2}$ at some high mass scale $M_F$, the charged fermions
will all acquire effective mass terms in the low-energy effective theory
via diagrams such as figure \ref{fig:fdiag}. The 
intermediate states are taken to be
vector-like fermions of mass $M = O(M_F)$, and we assume that the
fundamental couplings are $O(1)$. Figure \ref{fig:fdiag} then gives
an effective mass to the bottom quark,

\be
m_b \sim \vev{\phi_{WS}} \frac{\vev{W}}{M_F} \frac{\vev{\theta}^2}{M_F^2},
\ee
where $\vev{W}, \vev{\theta}$ are the vacuum expectation values
of Higgs fields $W$ and $\theta$ used to 
spontaneously break the $SMG \times U(1)^2$ down to the standard model.
The other charged fermions acquire their mass via similar diagrams.

As we discussed earlier we do not wish to extend the low-energy fermion
spectrum for reasons of minimality, so we have the usual SM fermion spectrum
with their usual representations under $SMG$. The fermion charges
under $U(1)_{f1}$ and $U(1)_{f2}$ are then severely constrained by the
requirement that all the anomalies involving them cancel. If we denote the
charges of the fermions under $U(1)_{f1}$ and $U(1)_{f2}$ by
$Q_{fi}(u_L) = u_{Li}$ ($i = 1, 2$) etc., then the anomaly constraints are
given by:
\bea
\mbox{Tr} [SU(3)^2 U(1)_{fi}] & = & 2(u_{Li} + c_{Li} + t_{Li})\nonumber \\ 
& & - (u_{Ri} + d_{Ri} + s_{Ri} + c_{Ri} + t_{Ri} + b_{Ri}) = 0, \nonumber\\ 
\Tr [SU(2)^2  U(1)_{fi}] & = & 3 (u_{Li} + c_{Li} + t_{Li})
+ e_{Li} + \mu_{Li} + \tau_{Li} = 0,\nonumber\\
\Tr [U(1)_Y^2 U(1)_{fi}] & = & u_{Li} + c_{Li} + t_{Li}
- 8 (u_{Ri} + c_{Ri} + t_{Ri}) \nonumber\\
& & -  2 (d_{Ri} + s_{Ri} + b_{Ri}) 
 + 3 (e_{Li} + \mu_{Li} + \tau_{Li}) \nonumber\\
& & - 6 (e_{Ri} + \mu_{Ri} + \tau_{Ri}) = 0, \nonumber\\
\Tr [U(1)_Y U(1)_{fi}^2] & = & u_{Li}^2 + c_{Li}^2 + t_{Li}^2 -
2 (u_{Ri}^2 + c_{Ri}^2 + t_{Ri}^2) \nonumber\\
& & + d_{Ri}^2 + s_{Ri}^2 + b_{Ri}^2 
 -(e_{Li}^2 + \mu_{Li}^2 + \tau_{Li}^2) \nonumber\\
& & + e_{Ri}^2 + \mu_{Ri}^2
+ \tau_{Ri}^2 = 0, \nonumber\\
%\Tr [U(1)^3_{f_i}] & = & 6 (u_{Li}^3 + c_{Li}^3 + t_{Li}^3)
%-3 (u_{Ri}^3 + c_{Ri}^3 + t_{Ri}^3 + d_{Ri}^3 + s_{Ri}^3 + b_{Ri}^3)
%\nonumber \\
%& & + 2 (e_{Li}^3 + \mu_{Li}^3 + \tau_{Li}^3) - (e_{Ri}^3 + \mu_{Ri}^3
%+ \tau_{Ri}^3) = 0,\nonumber\\
%\Tr[U(1)_{fi}^2 U(1)_{fj}] & = & 0, \nonumber\\
\Tr[U(1)_{fi} U(1)_{fj} U(1)_{fk}] & = & 6(u_{Li} u_{Lj} u_{Lk} +
c_{Li} c_{Lj} c_{Lk} + t_{Li} t_{Lj} t_{Lk}) \nonumber\\
& & -3(d_{Ri} d_{Rj} d_{Rk} +
s_{Ri} s_{Rj} s_{Rk} + b_{Ri} b_{Rj} b_{Rk}\nonumber\\
& & + u_{Ri} u_{Rj} u_{Rk} +
c_{Ri} c_{Rj} c_{Rk} + t_{Ri} t_{Rj} t_{Rk}) \nonumber\\
& & +2(e_{Li} e_{Lj} e_{Lk} +
\mu_{Li} \mu_{Lj} \mu_{Lk} + \tau_{Li} \tau_{Lj} \tau_{Lk})\nonumber\\
& & -(e_{Ri} e_{Rj} e_{Rk} + \mu_{Ri} \mu_{Rj} \mu_{Rk} + \tau_{Ri}
\tau_{Rj} \tau_{Rk})  = 0, \nonumber\\
\Tr[(\mbox{graviton})^2 U(1)_{fi}] & = & 6(u_{Li} + c_{Li} + t_{Li})
-3(u_{Ri} + d_{Ri} + s_{Ri} + c_{Ri} \nonumber\\
& & + t_{Ri} + b_{Ri}) 
 + 2 (e_{Li} + \mu_{Li} + \tau_{Li}) \nonumber\\
& & - (e_{Ri} + \mu_{Ri} + \tau_{Ri}) = 0.
\label{eq:anomaly}
\eea
\begin{table}[t]
\begin{displaymath}
\renewcommand{\arraystretch}{1.5}
\begin{array}{|c||*{11}{c|}}
\hline
 & \mbox{(1st. gen.)} & c_L & t_L &
c_R & s_R & t_R & b_R & \mu_L
& \tau_L & \mu_R & \tau_R \\ \hline \hline
Q_{f1} & 0 & 0 & 1 & 4 & 0 & 0 & -2 & 0 & -3 & 0 & -6 \\ \hline
Q_{f2} & 0 & -1 & 0 & -1 & 1 & -3 & 1 & 3 & 0 & 5 & 1\\ \hline
\end{array}
\end{displaymath}
\caption{An anomaly free choice of Abelian charges for
the fermion fields}
\label{tab:fcharges}
\end{table}
A possible choice of charges (which is based on the
AGUT Abelian charges) satisfying these constraints is given
in table \ref{tab:fcharges} and, as we shall see, a realistic charged
fermion mass spectrum can be obtained for these charges by making
a suitable choice of Higgs fields. The set of charges in table
\ref{tab:fcharges} is not the only one which is anomaly free.
For example, the AGUT has four
$U(1)$s, with linearly independent sets of charges which satisfy 
the anomaly constraints of eq.
\ref{eq:anomaly}.  
In the AGUT (\cite{AGUT1}) one of these $U(1)$s is broken before the others
at the Planck scale, leaving three unbroken $U(1)$ generators.
In this paper we choose the fermion charges to be a linear
combination 
of the charges under these unbroken generators. 
(Our choice of charges is given by $Q_Y = y_1 + y_2 + y_3,
Q_{f1} = 3 y_3$ and $Q_{f2} = -3 y_2 + Q_f$
where $y_{1, 2, 3}$ and $Q_f$ are the AGUT fermion charges of
reference \cite{AGUT1}).
We could alternatively have chosen to use the charges under the broken
$U(1)$ for $Q_{f1}$ or $Q_{f2}$; however, we are unaware
of any choice of charges (with only two non-standard model
$U(1)$s) involving this broken $U(1)$ which yields
a realistic charged fermion spectrum.

The Weinberg-Salam Higgs field, $\phi_{WS}$, charges are chosen so
that the top quark obtains its mass directly from its Yukawa coupling with
$\phi_{WS}$, and $M_t$ is thus unsuppressed. The other fermions cannot couple
directly to $\phi_{WS}$ since such couplings are protected by
the chiral symmetries. Hence we introduce three other Higgs fields
$W$, $\xi$ and $\theta$ to break the $U(1)_{f1}$ and $U(1)_{f2}$ 
with charges and vacuum expectation values chosen to 
give a realistic fermion spectrum. The charges and vacuum expectation
values of the Higgs fields  are given in table \ref{tab:hcharges}.
We take the Higgs fields at the fundamental scale
to be singlets under the standard model symmetries.
The charged fermion effective SM Yukawa matrices are then given by
\bea
H_U & \sim & \left ( \begin{array}{ccc}
        \vev{W}\vev{\theta}^4\vev{\xi}^2 &
                \vev{W}^2 \vev{\theta}^2 \vev{\xi} & \vev{W}\vev{\theta}^4
		\vev{\xi}\\
        \vev{W}\vev{\theta}^4\vev{\xi}^3 & 
                \vev{W}^2\vev{\theta}^2 & \vev{W}\vev{\theta}^4 \\
        \vev{\xi}^3  & \vev{W}\vev{\theta}^2 & 1
                        \end{array} \right ), \label{H_U} \\
H_D & \sim & \left ( \begin{array}{ccc}
        \vev{W}\vev{\theta}^4\vev{\xi}^2 & \vev{W}\vev{\theta}^4\vev{\xi} & 
	\vev{\theta}^6\vev{\xi} \\
        \vev{W}\vev{\theta}^4\vev{\xi} & \vev{W}\vev{\theta}^4 & 
	\vev{\theta}^6 \\
        \vev{W}^2\vev{\theta}^8\vev{\xi} & \vev{W}^2\vev{\theta}^8 & \vev{W}
	\vev{\theta}^2 \end{array} \right ), \label{H_D} \\
H_E & \sim & \left ( \begin{array}{ccc}
        \vev{W}\vev{\theta}^4\vev{\xi}^2 & \vev{W}\vev{\theta}^4\vev{\xi}^3 &
                \vev{W}\vev{\theta}^8\vev{\xi} \\
       \vev{W}\vev{\theta}^4\vev{\xi}^5 & \vev{W}\vev{\theta}^4 & 
        \vev{W}\vev{\theta}^8\vev{\xi}^2 \\
        \vev{W}\vev{\theta}^{10}\vev{\xi}^3 & \vev{W}^2\vev{\theta}^8 & 
	\vev{W}\vev{\theta}^2
                        \end{array} \right ) \label{H_E},
\end{eqnarray}
where the Higgs field vacuum expectation values $\vev{W},
\vev{\xi}$ and $\vev{\theta}$ are in units 
of the fundamental scale, $M_F$.
These mass matrices yield exactly the same masses and mixings at the
fundamental scale as
we obtained in the AGUT model in previous papers \cite{AGUT1}, as can
be seen by substituting the Higgs field combination $\theta^2$ in this paper
by the Higgs field $T$ in the AGUT, and relabelling the $c_R$ and $t_R$
fields. 
\begin{table}[t]
\begin{displaymath}
\renewcommand{\arraystretch}{1.5}
\begin{array}{|c||c|c|c|c|}\hline
& y/2 & Q_{f1} & Q_{f2} &
\mbox{Vacuum expectation value}\\ 
\hline \hline
\phi_{WS} & \frac{1}{2} & -1 & -3 & \\ \hline
W & 0 & 3 & \frac{5}{3} & 0.158\\ \hline
\theta & 0 & \frac{1}{2} & \frac{1}{6} & 0.266 \\ \hline
\xi & 0 & 0 & 1 & 0.099\\ \hline 
\end{array}
\end{displaymath}
\caption{Higgs field charges which have been chosen to give a realistic
charged fermion spectrum, and the vacuum expectation values for the
chiral symmetry breaking Higgs fields in units of the fundamental
scale $M_F$.}
\label{tab:hcharges}
\end{table}
This is because
(after this trivial relabelling of fermion fields) 
the charges on the fermion fields are
the same as a linear combination of the remaining Abelian fermion charges
in the AGUT after one of the AGUT $U(1)$'s is spontaneously broken. The choice
of Higgs fields in the $SMG \times U(1)^2$ model is however different
and, whilst this leads to the same charged fermion spectrum as in the
AGUT (see table \ref{best-fit} for the best fit spectrum from \cite{AGUT1}), 
it does not yield the same neutrino spectrum. The AGUT cannot produce
the same neutrino mass matrix structure (without increasing the
number of Higgs fields), since it is not possible
to choose a consistent set of non-Abelian representations for the Higgs fields.
We shall see however  that, within the $SMG \times U(1)^2$ model,
we can obtain an acceptable neutrino spectrum.
\begin{table}
\caption{Best fit to conventional experimental data.
All masses are running
masses at 1 GeV except the top quark mass which is the pole mass.}
\begin{displaymath}
\begin{array}{ccc}
\hline
 & \mathrm{Fitted} & \mathrm{Experimental} \\ \hline
m_u & 3.6 \mathrm{\; MeV} & 4 \mathrm{\; MeV} \\
m_d & 7.0 \mathrm{\; MeV} & 9 \mathrm{\; MeV} \\
m_e & 0.87 \mathrm{\; MeV} & 0.5 \mathrm{\; MeV} \\
m_c & 1.02 \mathrm{\; GeV} & 1.4 \mathrm{\; GeV} \\
m_s & 400 \mathrm{\; MeV} & 200 \mathrm{\; MeV} \\
m_{\mu} & 88 \mathrm{\; MeV} & 105 \mathrm{\; MeV} \\
M_t & 192 \mathrm{\; GeV} & 180 \mathrm{\; GeV} \\
m_b & 8.3 \mathrm{\; GeV} & 6.3 \mathrm{\; GeV} \\
m_{\tau} & 1.27 \mathrm{\; GeV} & 1.78 \mathrm{\; GeV} \\
V_{us} & 0.18 & 0.22 \\
V_{cb} & 0.018 & 0.041 \\
V_{ub} & 0.0039 & 0.0035 \\ \hline
\end{array}
\end{displaymath}
\label{best-fit}
\end{table}

\section{Neutrino masses and mixings from an explicit model}
\label{sec:neutrino}

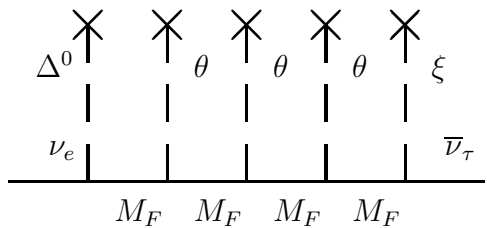
\begin{figure}
\begin{picture}(40000,3000)
\THICKLINES
\drawline\fermion[\E\REG](8000,1000)[3000]
\global\advance \pmidy by 1000
\put(\pmidx,\pmidy){$\nu_e$} 
\drawline\fermion[\E\REG](11000,1000)[3000]
\global\advance \pmidy by -1500
\global\advance \pmidx by -500
\put(\pmidx,\pmidy){$M_F$}

\drawline\fermion[\E\REG](14000,1000)[3000]
\global\advance \pmidy by -1500
\global\advance \pmidx by -500
\put(\pmidx,\pmidy){$M_F$}

\drawline\fermion[\E\REG](17000,1000)[3000]
\global\advance \pmidy by -1500
\global\advance \pmidx by -500
\put(\pmidx,\pmidy){$M_F$}

\drawline\fermion[\E\REG](20000,1000)[3000]
\global\advance \pmidy by -1500
\global\advance \pmidx by -500
\put(\pmidx,\pmidy){$M_F$}

\drawline\fermion[\E\REG](23000,1000)[3000]
\global\advance \pmidy by 1000
\put(\pmidx,\pmidy){$\overline{\nu}_{\tau}$}

\drawline\scalar[\N\REG](11000,1000)[3]
\global\advance \pmidx by -2000
\global\advance \pmidy by 1000
\put(\pmidx,\pmidy){$\Delta^0$}
\global\advance \scalarbackx by -530
\global\advance \scalarbacky by -530
\drawline\fermion[\NE\REG](\scalarbackx,\scalarbacky)[1500]
\global\advance \scalarbacky by 1060
\drawline\fermion[\SE\REG](\scalarbackx,\scalarbacky)[1500]

\drawline\scalar[\N\REG](14000,1000)[3]
\global\advance \pmidx by 1000
\global\advance \pmidy by 1000
\put(\pmidx,\pmidy){$\theta$}
\global\advance \scalarbackx by -530
\global\advance \scalarbacky by -530
\drawline\fermion[\NE\REG](\scalarbackx,\scalarbacky)[1500]
\global\advance \scalarbacky by 1060
\drawline\fermion[\SE\REG](\scalarbackx,\scalarbacky)[1500]

\drawline\scalar[\N\REG](17000,1000)[3]
\global\advance \pmidx by 1000
\global\advance \pmidy by 1000
\put(\pmidx,\pmidy){$\theta$}
\global\advance \scalarbackx by -530
\global\advance \scalarbacky by -530
\drawline\fermion[\NE\REG](\scalarbackx,\scalarbacky)[1500]
\global\advance \scalarbacky by 1060
\drawline\fermion[\SE\REG](\scalarbackx,\scalarbacky)[1500]

\drawline\scalar[\N\REG](20000,1000)[3]
\global\advance \pmidx by 1000
\global\advance \pmidy by 1000
\put(\pmidx,\pmidy){$\theta$}
\global\advance \scalarbackx by -530
\global\advance \scalarbacky by -530
\drawline\fermion[\NE\REG](\scalarbackx,\scalarbacky)[1500]
\global\advance \scalarbacky by 1060
\drawline\fermion[\SE\REG](\scalarbackx,\scalarbacky)[1500]

\drawline\scalar[\N\REG](23000,1000)[3]
\global\advance \pmidx by 1000
\global\advance \pmidy by 1000
\put(\pmidx,\pmidy){$\xi$}
\global\advance \scalarbackx by -530
\global\advance \scalarbacky by -530
\drawline\fermion[\NE\REG](\scalarbackx,\scalarbacky)[1500]
\global\advance \scalarbacky by 1060
\drawline\fermion[\SE\REG](\scalarbackx,\scalarbacky)[1500]
\end{picture}
\vskip .3cm
\caption{Example Feynman diagram for neutrino mass in the $SMG \times
U(1)^2$ model.}
\label{fig:nmass}
\end{figure}

Neutrino masses can be generated in this model by the Weinberg-Salam
Higgs field, via a see-saw like mechanism, giving a dominant
off-diagonal element in the neutrino mass matrix, 
\be
M_{\nu} \sim \frac{\phi^2_{WS}}{M_F}\left ( \begin{array}{ccc}
        \vev{W}^2\vev{\theta}^8\vev{\xi}^4 &
                \vev{W}^2\vev{\theta}^8\vev{\xi} &
                \vev{W}^2\vev{\theta}\vev{\xi}^3 \\
        \vev{W}^2\vev{\theta}^8\vev{\xi} &
                \vev{W}\vev{\theta}^{10} & \vev{W}^2\vev{\theta} \\
        \vev{W}^2\vev{\theta}\vev{\xi}^3 & \vev{W}^2\vev{\theta} &
                \vev{W}^2\vev{\theta}^2\vev{\xi}^2
                        \end{array} \right ).
\label{hnu}
\ee
This yields
nearly maximal $\nu_{\mu} - \nu_{\tau}$ mixing 
between a nearly degenerate pair of neutrinos. As we discussed earlier
this does not lead to an acceptable phenomenology,
and hence we require
a different mechanism to generate the dominant contribution
to the neutrino masses and mixings.
We do this here by introducing an $SU(2)$ triplet Higgs field
$\Delta$. The charges on this Higgs field are then chosen so that the
(1, 2) and (1, 3) elements of $M_{\nu}$
are suppressed by equal amounts, giving
\be
\left(\frac{y}{2}, Q_{f1}, Q_{f2}\right) = \left(1, \frac{3}{2}, -\frac{3}{2}
\right).
\ee
The neutrino mass matrix,
\be
M_{\nu} \sim \vev{\Delta^0} \vev{\theta}^3\left( \begin{array}{ccc}
\vev{\xi}^2 & \vev{\xi} & \vev{\xi}\\
\vev{\xi} & \vev{\xi}^4 & \vev{\xi}^2\\
\vev{\xi} & \vev{\xi}^2 & \vev{\theta}^6
\end{array} \right),
\ee
is then generated by diagrams such as figure \ref{fig:nmass}.
We have ignored CP violating phases here, and there are unknown
$O(1)$ factors in front of each of the mass matrix elements.

This mass matrix gives
\be
\frac{\Delta m^2_{12}}{\Delta m^2_{23}} \sim \vev{\xi}
\ee
which is not small enough for the `just-so' or MSW solutions to the
solar neutrino problem if we take
\be
\Delta m^2_{23} \sim 10^{-3} \eV^2
\ee
for the atmospheric neutrino problem. Hence we shall use the
`energy-independent' vacuum oscillation solution to the solar
neutrino problem. The mixing from this mass matrix is similar
to that given by eq. \ref{eq:Unu}, although the elements
of order $\vev{\Delta^0}\theta^3 \xi^2$ in the mass matrix can have some
effect on the mixing leading to some small deviations
from the form of eq. \ref{eq:Unu}. The electron
neutrino mixing remains very close to maximal regardless of the
$O(1)$ factors in the mass matrix, and makes almost
no contribution to the atmospheric neutrino mixing. Depending on the $O(1)$
factors the muon and tau neutrino mixing can differ slightly from
that given by eq. \ref{eq:Unu},
although if eq. \ref{eq:b/a} is satisfied then the mixing between
them remains large enough to solve the atmospheric neutrino problem.

Hence if we take $\vev{\Delta} \sim 12 \eV$ to give suitable masses
for the atmospheric neutrino problem then we have
\bea
\Delta m^2_{12} & \sim & 10^{-4} \eV^2, \, \sin^2 2\theta_{12}
\sim 1 \\
\Delta m^2_{23} & \sim & 10^{-3} \eV^2, \, \sin^2 2\theta_{23}
= 0.7 - 1.0
\eea
for the solar and atmospheric neutrinos respectively. This means we
will have an electron neutrino flux suppression of $1/2$ for all of
the solar neutrinos, and the atmospheric neutrino problem will
be due to large $\nu_{\mu} - \nu_{\tau}$ mixing.
The neutrino masses are too small to make
a significant contribution to dark matter, or to the anomaly observed at LSND
\cite{LSND}.
Hence we predict that the LSND result  will prove to be unfounded.
The amplitude of neutrinoless double beta decay is proportional
to $(M_{\nu})_{ee}$, which we predict to be $(M_{\nu})_{ee} \sim
2 \times 10^{-3} \eV$, which is much less than the current limit
of $(M_{\nu})_{ee} \le 0.45 \eV$ \cite{beta} and the sensitivities of
current or planned experiments.

In obtaining the spectrum of neutrino masses we have simply
chosen $\vev{\Delta^0}$ to have the required value for the
atmospheric neutrinos. However, there is some unnaturalness in
obtaining a suitable value for $\vev{\Delta^0}$ from the scalar
potential. If we write down the low energy effective
scalar potential we have
\bea
V(\phi_{WS}, \Delta) & \sim & \lambda \{ (\phi_{WS}^{\dagger}
\phi_{WS})^2 + \lambda^{\prime} (\Delta^{\dagger} \Delta)^2
+ \lambda^{\prime \prime} M_F \phi_{WS}^{\dagger2} \Delta \vev{W}
\vev{\xi}^2 \vev{\theta} \nonumber \\
& & - \eta M_F^2 \Delta^{\dagger}
\Delta - \frac{\mu^2}{\lambda} \phi_{WS}^{\dagger} \phi_{WS}\}
\eea
where we would typically expect $\lambda^{\prime}, \lambda^{\prime
\prime}, \eta = O(1).$ However, this leads to a vacuum
expectation value for $\Delta$ of
\be
\vev{\Delta^0} \sim \frac{\vev{\phi_{WS}^2}}{M_F} 
\vev{W} \vev{\xi}^2 \vev{\theta}.
\ee
Whilst we can choose $M_F$ to give the required vacuum expectation
value for $\Delta$ we then find that, since
$\vev{\Delta}$ is much less than the see-saw
scale $\frac{\vev{\phi_{WS}}^2}{M_F}$, the neutrino mass
matrix is dominated by the see-saw type diagrams which as we noted
earlier, do not yield an acceptable phenomenology. Hence, in order
to avoid this problem, we would require a $\phi_{WS}^{\dagger2} \Delta$
coupling which is for some unknown reason much larger than expected.
Of course the scalar potential is in any case not well understood,
since the lightness of the Weinberg-Salam Higgs field is also something
of a mystery.

It should be noted that, whilst in this case we have some difficulty
in obtaining a suitable vacuum expectation value for the triplet
Higgs field, this will not necessarily be the case for other models
which use this mechanism for generating the neutrino masses. If the 
see-saw neutrino masses are sufficiently suppressed by the symmetry breaking
parameters, then the masses coming from the triplet Higgs field will
dominate and there will be no problem.

\section{Conclusions}

We have shown that models with only the 3 standard model neutrinos 
(in the low energy spectrum), and chiral symmetry
breaking can explain the solar and atmospheric neutrino problems
including the Super-Kamiokande zenith angle distribution. This can occur
if the chiral symmetry does not lead to (independent)
elements in $M_{\nu}$ which are all of different orders
of magnitude (as we assumed in a previous paper). The atmospheric
neutrino problem is explained by large $\nu_{\mu} - \nu_{\tau}$ mixing,
and (for the mass matrix structure we examined) the solar neutrino
deficit is due to nearly maximal electron neutrino vacuum oscillations,
which can be either `just-so' or `energy-independent'.  We presented
an explicit model, which is an anomaly free Abelian extension of the SM,
yielding this type of phenomenology, although there are unresolved
problems in the scalar potential.
This model is an extension
of a model which gives a realistic 3 parameter fit to
 the charged fermion masses and mixings.
It gives an `energy-independent' solar neutrino suppression
of $1/2$, with $\Delta m^2_{solar} \sim 10^{-4} \eV^2$. We
also predict that the signal at LSND will not be confirmed
by other experiments, and that
the neutrinos will not make a significant contribution to hot dark matter.

The prospects for examining this scenario are good. Experiments
such as SNO \cite{SNO}, Borexino \cite{Borexino} and KamLand \cite{Kamland} 
should provide us with more information
on the solar neutrino spectrum. 
Super-Kamiokande
will also provide data on the day-night asymmetry and seasonal variations
which will be important in determining the type of solution to the
solar neutrino problem. Long baseline experiments such as K2K \cite{K2K}
and MINOS \cite{MINOS}
should enable us to confirm the nature of the atmospheric neutrino
oscillations with a better understood neutrino source, and should 
tell us whether the $\nu_{\mu}$ oscillations are to $\nu_{\tau}$
or a sterile neutrino. The LSND
result will also be further tested by Karmen at $95 \% \,C.L.$, and
definitively by MiniBoone; neither of which we would expect
to find evidence of oscillations. In conclusion, we predict the atmospheric
neutrino problem to be due to large $\nu_{\mu} - \nu_{\tau}$ oscillations
with $\Delta m^2 \sim 10^{-3} \eV^2$, and the solar neutrino
deficit to be due to electron neutrino vacuum oscillations of either
the `just-so' or `energy-independent' type. This scenario should
be confirmed or denied by a number of experiments in the near future.

\section*{Acknowledgements}

H.B.N. and
C.F. acknowledge funding from INTAS 93-3316-ext, and the EU
grant HMC 94-0621. M.G. is grateful for a PPARC studentship.
We would also like to thank M. Jezabek for 
useful discussions.

\end{document}